\documentclass[aps,prl,twocolumn,superscriptaddress,amsmath,amssymb]{revtex4}
\usepackage{graphicx}
\usepackage{amsmath}
\usepackage{pstricks}
\usepackage{tikz}
\usepackage{float}

\begin{document}

\title{Curvature Induced Topological Defects of $p$-wave Superfluid on a Sphere}
\author{Ruihua Fan}
\affiliation{Institute for Advanced Study, Tsinghua University, Beijing, 100084, China}
\affiliation{Department of Physics, Peking University, Beijing, 100871, China}
\author{Pengfei Zhang}
\affiliation{Institute for Advanced Study, Tsinghua University, Beijing, 100084, China}
\author{Zhe-Yu Shi}
\email{zheyu.shi@monash.edu}
\affiliation{Institute for Advanced Study, Tsinghua University, Beijing, 100084, China}
\affiliation{School of Physics and Astronomy, Monash University, Victoria 3800, Australia}
\date{\today}

\begin{abstract}
We study the ground state of spinless fermions living on a sphere across $p$-wave Feschbach resonances.
%In this letter, we first established a microscopic model describing p-wave superfluid on any curved surface and studied the sphere with great care.
By construsting a microscopic model of fermions on a general curved surface, we show that
the Guassian curvature induces an emergent magnetic field coupled to the $p\pm ip$ order parameters. In the case of a sphere, the magnetic field corresponds to a Dirac monopole field, which causes topological defects in the superfluid ground state. Using the BCS mean field theory, we calculate its many-body ground state self consistently and give the phase diagram. The ground state may exhibit two types of topological defects, two voritces on the south and north pole or a domain wall which separates  $p_\theta+ ip_\phi$ and $p_\theta-ip_\phi$ superfluids.
%By self-consistent numerics, we found that when the system goes from BEC to BCS limit, the change of ratio between Cooper pair size and healing length will lead to a change from $p+ip$ phase to $p_\phi$ phase. And within the domain wall configurations, there is a $Z_2$ phase transition due to the competition between the kinetic energy of the two chiral components of superfluid.
\end{abstract}

\maketitle
{\it Introduction.}
%The discovery of BEC-BCS crossover in atomic gases with $s$-wave interactions is a milestone in cold atom physics\cite{BEC-BCS1,BEC-BCS2}. Recent development in the $p$-wave Feschbach resonances of fermonic atoms has paved the way to the realization of $p$-wave superfluidity.
Recent development of the $p$-wave Feschbach resonance technique in ultracold fermonic atoms\cite{p-wave exp1,p-wave exp2,p-wave exp3,p-wave exp4,p-wave exp5} has opened up possibilities to many novel quantum states. Among these states, the chiral $p$-wave superfluid\cite{Andreev,Gurarie,Radzihovsky,p-wave mean field Yip,p-wave crossover Pf Zhuang} has attracted significant amount of interest, not only because it is a nontrivial topological phase of matter, but also because the system may provide a platform for fault-tolerant quantum computation\cite{q computation1, q computation2, q computation3}.

On the other hand, more and more newly developed trapping techniques allow the experimentalists to confine cold atomic gases on various extraordinary geometries such as helices\cite{fiber1,fiber2} and spheres\cite{Pfau1,Pfau2}. This provides a unique opportunity to explore quantum many-body systems in non-Euclidean spaces and study the effects of the spatial curvatures on these systems.

In this manuscript, we consider a system of spinless fermions living on a sphere interacting through a $p$-wave contact potential. Previous studies of such system on a flat plane shows that the ground state is a $p_x\pm ip_y$ paired superfluid with uniform order parameter in space\cite{Andreev,Gurarie,Radzihovsky,p-wave mean field Yip,p-wave crossover Pf Zhuang}. 
If the system lives on a sphere, it has been known that the system must support topological defects\cite{Read,Moroz}. In ref.\cite {Moroz}, the authors argued based on the nature of the chiral superfluid and used the Poincar\'{e}-Hopf index theorem to reach this conclusion. In this work, we demonstrate it from a few-body point of view. We use a two-body calculation to show that the two-body bound state with nonzero angular momentum would experience an emergent monopole field when moving on the sphere\cite{2body}. And the magnet flux of the monopole necessarily cause topological defects such as vortices and domain walls on the sphere.
\begin{figure}[h]\centering
	\includegraphics[height=3.2in]
	{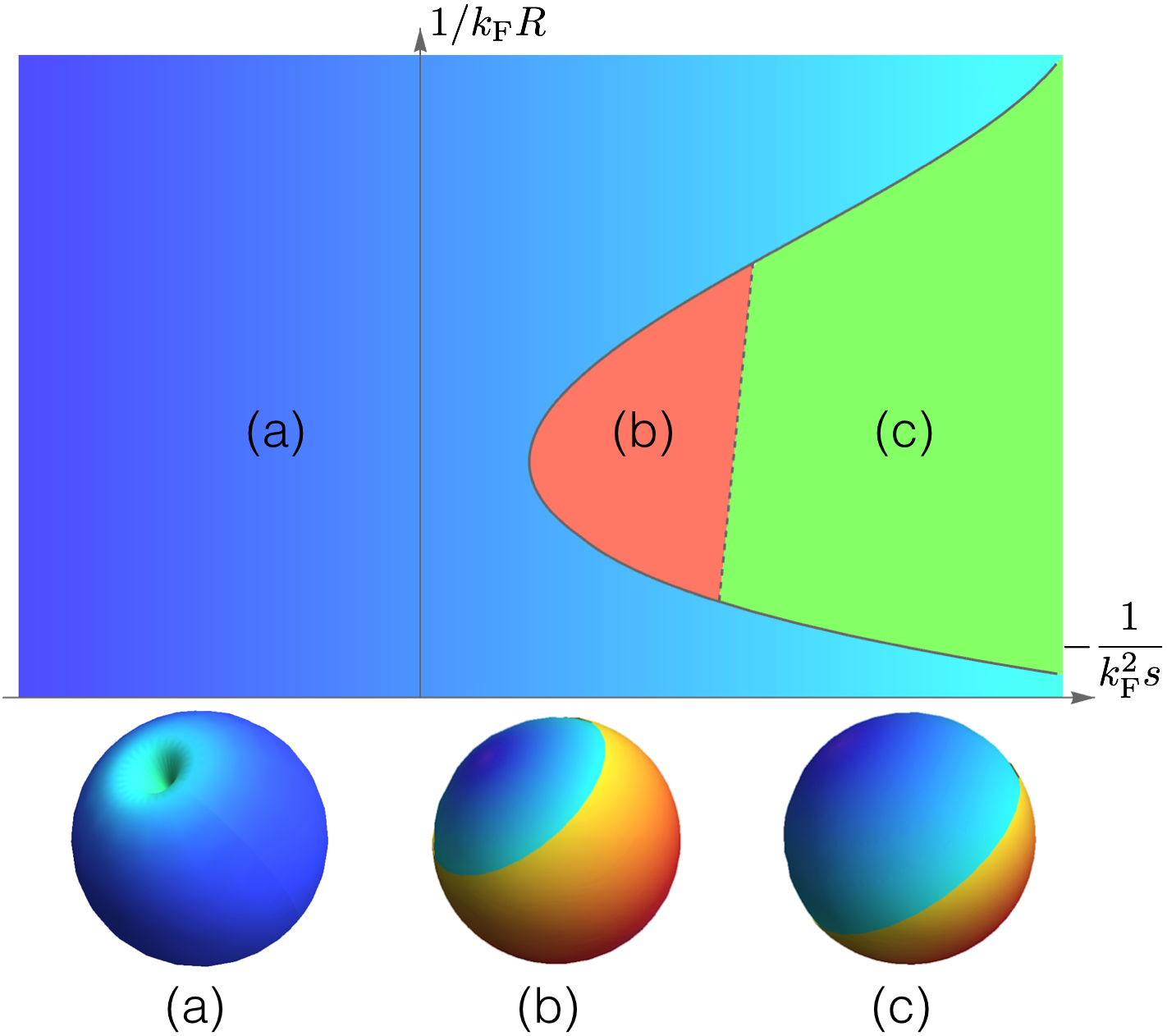}
	\caption{A schematic plot of the phase diagram. Warm colors denotes $p+ip$ order parameter. Cold colors denotes $p-ip$ order parameter. $R$ is the radius of the sphere, $s$ is the low energy $p$-wave scattering area. Three phases corresponding to different configurations of topological defects appear, (a): a vortex pair on the south and north poles. (b): a domain wall on the southern or northern hemisphere. (c): a domain wall on the equator.}\label{phase_d}
\end{figure}

Our main results are summarized in the ground state phase diagram fig.(\ref{phase_d}). The ground state may present three different phases depending on the radius of the sphere $R$ and the interaction strength indicated by scattering area $s$. (a) phase corresponds to the configuration of a pair of vortices located on the south and north pole of the sphere separately. While both (b) and (c) phase correspond to a solution with a domain wall between the $p_\theta+ip_\phi$ and $p_\theta-ip_\phi$ superfluids. The difference is that in (b) phase the domain wall is close to one pole and locates on the southern or northern hemisphere, and in (a) phase it is always on the equator.

{\it Model.} 
While most of the previous studies treated this subject phenomenologically\cite{curved space field theory Son,curved space field theory Hoyos}, in this work, we target this problem from a {\it microscopic} model. We begin with a Hamiltonian describing spinless fermions with $p$-wave interaction on a general two-dimensional surface $\mathcal{M}$:

\begin{eqnarray}
	H=\int_\mathcal{M} \mathrm{d}s\  \mathcal{H},
\end{eqnarray}
\begin{eqnarray}
	\mathcal{H}&=&\psi^\dagger\bigg{(}-\frac{D^\nu D_\nu}{2m_f}-\mu\bigg{)}\psi+d_\alpha^\dagger\bigg{(}-\frac{D^\nu D_\nu}{4m_f}+\epsilon-2\mu\bigg{)}d^\alpha\nonumber\\
	&&-i\lambda(d_\nu\psi^\dagger D^\nu\psi^\dagger+d_\nu^\dagger\psi D^\nu\psi).\label{H_density}
\end{eqnarray}
Here $\mathrm{d}s=\sqrt{g}dx^1dx^2$ is the area element of the surface with $g=\det g_{\mu\nu}$. $\psi$($d_\nu$) is the annihilation operator for the open channel fermions(closed channel molecules). $D_\nu$($D^\nu$) stands for the covariant(contravariant) derivative with respect to the $\nu$ th coordinate. We use the Einstein summation convention and all the Greek letters range over $\{1,2\}$.

Note that if we set $g_{\mu\nu}=\delta_{\mu\nu}$(planar case), the Hamiltonian becomes the well studied two-channel model for the 2D $p$-wave superfluid\cite{Andreev, Radzihovsky, Gurarie}. While for general curved surfaces, our model shows several remarkable features because of the nontrivial metric $g_{\mu\nu}$.

First of all, to preserve the covariance of the Hamiltonian we introduce both the covariant and contravariant molecular operators $d_{\nu}$ and $d^\nu$, which are related by $d_{\mu}=g_{\mu\nu}d^\nu$. 
	
Secondly, to guarantee the particle number conservation, the commutation relations for both fields become
\begin{eqnarray}
\{\psi(x),\psi^\dagger(y)\}=\frac{1}{\sqrt{g}}\delta^{(2)}(x-y),\\
{}[d^\mu(x),d_\nu(y)]=\frac{1}{\sqrt{g}}\delta^\mu_\nu\delta^{(2)}(x-y).
\end{eqnarray}
Given these relations, it is straightforward to define the conserved particle number operator,
\begin{eqnarray}
N=N_F+2N_B=\int_{\mathcal{M}} \mathrm{d}s \bigg{(}\psi^\dagger\psi+2d_\nu^\dagger d^\nu\bigg{)}.
\end{eqnarray}

Moreover, the covariant and contravariant derivative in the Hamiltonian lead to an extra term due to the connection of the surface.
Take the unit sphere (radius $R=1$) as an example. We shall use the usual spherical coordinates $x^1=\theta,\ x^2=\phi$. The fermion kinetic energy term can be computed,
\begin{eqnarray}
	\psi^\dagger{D^\nu D_\nu}\psi=-\psi^\dagger{L^2}\psi,
\end{eqnarray}
where $L^2=-(\frac{1}{\sin\theta}\frac{\partial}{\partial\theta}(\sin\theta\frac{\partial}{\partial\theta})+\frac{1}{\sin^2\theta}\frac{\partial^2}{\partial\phi^2})$ is the total angular momentum operator.

While for the molecular field, since the operator $D^\nu D_\nu$ is acting on a contravariant vector field $d^\alpha$, the derivative will give an extra connection term. Consequently, it cannot be simply replaced by $L^2$ operator. Inspired by the two-body calculation\cite{2body}, we recombine $d^{\alpha}$ operators and define $d_\pm^\dagger=\frac{1}{\sqrt{2}}(d^{1\dagger}\pm i\sin\theta d^{2\dagger})$, which represents the close channel molecules with different chiralities(angular momentums). Using these two operators, the kinetic energy can be expressed in a simple form,
\begin{eqnarray}
	d_\alpha^\dagger D^\nu D_\nu d^\alpha=-d_\pm^\dagger{(\mathbf{L}\pm \hat{r}\times \mathbf{A})^2}d_\pm
\end{eqnarray}
with $\mathbf{A}=-\cot \theta \hat{e}_\phi$.

Besides the common $L^2$ term, there emerges an extra gauge field $\mathbf{A}$ for the molecular field, and the {\it effective charges} for molecules with different chiralities are opposite. The corresponding magnetic field $\nabla\times\mathbf{A}=\hat{e}_r/R^2$ is exactly the magnetic field of a Dirac monopole with charge $1$. Consequently, the superfluid ground state necessarily supports topological defects such as vortices and domain walls.

{\it Two-body problem.} We first consider a two-fermion problem on the sphere, which helps to relate the bare parameters $\epsilon$ and $\lambda$ to physical parameters. 
A two-body state can be generally written as
\begin{eqnarray}
	|\psi\rangle=\left[\int\varphi(\mathbf{r}_1,\mathbf{r}_2)\psi^\dagger(\mathbf{r}_1)\psi^\dagger(\mathbf{r}_2)+\int\chi_\pm(\mathbf{r})d_\pm^{\dagger}(\mathbf{r})
	\right]
	|0\rangle,
\end{eqnarray}
where $\varphi(\mathbf{r}_1,\mathbf{r}_2)$ is the two-body wave function and $\chi_\pm(\mathbf{r})$ is the dimer wave function.

Substituting it into $(E-H)|\psi\rangle=0$, we obtain the Schr\"{o}dinger equation for $\varphi(\mathbf{r}_1,\mathbf{r}_2)$ and $\chi_\pm(\mathbf{r})$,
\begin{eqnarray}
	\bigg{(}E-\frac{L_1^2}{2m_f}-\frac{L_2^2}{2m_f}\bigg{)}\varphi=-i\lambda\chi_\pm(\mathbf{r}_c)D_\pm\delta^{(2)}(\mathbf{r}_{12}),\label{eq_varphi}\\\bigg{(}E-\frac{(\mathbf{L}\pm\hat{r}\times \mathbf{A})^2}{4m_f}-\epsilon\bigg{)}\chi_\pm=i2\lambda D_\pm\varphi\big{|}_{\mathbf{r}_{12}=0}.\label{eq_chi}
\end{eqnarray}
where $\mathbf{r}_c=\frac{\mathbf{r}_1+\mathbf{r}_2}{2}$ and $\mathbf{r}_{12}=\mathbf{r}_1-\mathbf{r}_2$ are the center-of-mass and relative coordinates, and
$D_\pm$ is defined as $D_\pm=\frac{1}{\sqrt{2}}(D^1\pm i\sin\theta D^2)$.

The term with $\delta$-function on the R.H.S. of Eq. (\ref{eq_varphi}) may be regarded as a contact pseudopotential\cite{pseudop1,pseudop2}, which can be replaced by a short-range Bethe-Peierls boundary condition. This helps to relate $\epsilon$ and $\lambda$ to physical parameters such as scattering area $s$ and effective range $r_0$. Both $s$ and $r_0$ describe the low energy scattering properties of the $p$-wave interaction\cite{QDT, Peng Zhang, Braateen}, which lead to universal physics for many-body system\cite{p-wave contact 3D zhenhua, p-wave contact 3D Ueda, contact spectral 3D Qi,p-wave contact 2D,p-wave 3-body contact 2D}. It can be shown that,
\begin{eqnarray}
	\frac{1}{\lambda^2}=-\frac{m_f^2}{2\pi}\log\Lambda r_0,\quad\frac{\epsilon}{\lambda^2}=-\frac{m_f}{4s}+\frac{m_f}{4\pi}\Lambda^2,\label{r_relation}
\end{eqnarray}
where $\Lambda$ is the momentum cutoff.

Note that the physical parameters and the renormalization relation (\ref{r_relation}) take the exactly same form as the two-body problem on a 2D plane\cite{2body,Andreev}, which reflects the short-range nature of the interaction.

Detailed results of the two-body problem can be found in ref.{\cite{2body}}. Here we make a few remarks on the interesting many-body effects from the two-body consideration.

i). As mentioned previously, the emergence of magnetic field $\mathbf{A}$ will significantly change the ground state configurations of order parameter. Especially in the $s\rightarrow+\infty$(BEC) limit, where two fermions can form a deeply bound state, the many-body system can be viewed as a cloud of weakly interacting bosonic molecules. And the bosons should condense to the lowest Landau level on the sphere. In the monopole field $\mathbf{A}$, the lowest Landau level is three-fold degenerate and the eigen functions(monopole harmonics) are given by Wigner-D matrices $D^{1*}_{m,\pm 1}$\cite{explicit exp}. This also serves as a benchmark for our many-body calculation.

ii). From Eq.(\ref{eq_chi}), one can see that two dimer wave functions $\chi_+$ and $\chi_-$ are coupled to each other through two-body wave function $\varphi$. The coupling corresponds to the physical process that a $d_+$ molecule breaks into two fermions and recombines into a $d_-$ molecule on the sphere. This transition is a unique feature of curved surfaces because the angular momentum conservation would forbid the transition process on the plane. Thus, on the many-body level, we may expect the $p_\theta+ip_\phi$ Cooper pairs can also be transformed into the $p_\theta-ip_\phi$ Cooper pairs due to the curvature effect, which can enrich possible phases of the ground state.

\begin{figure*}
	\includegraphics[height=2in]{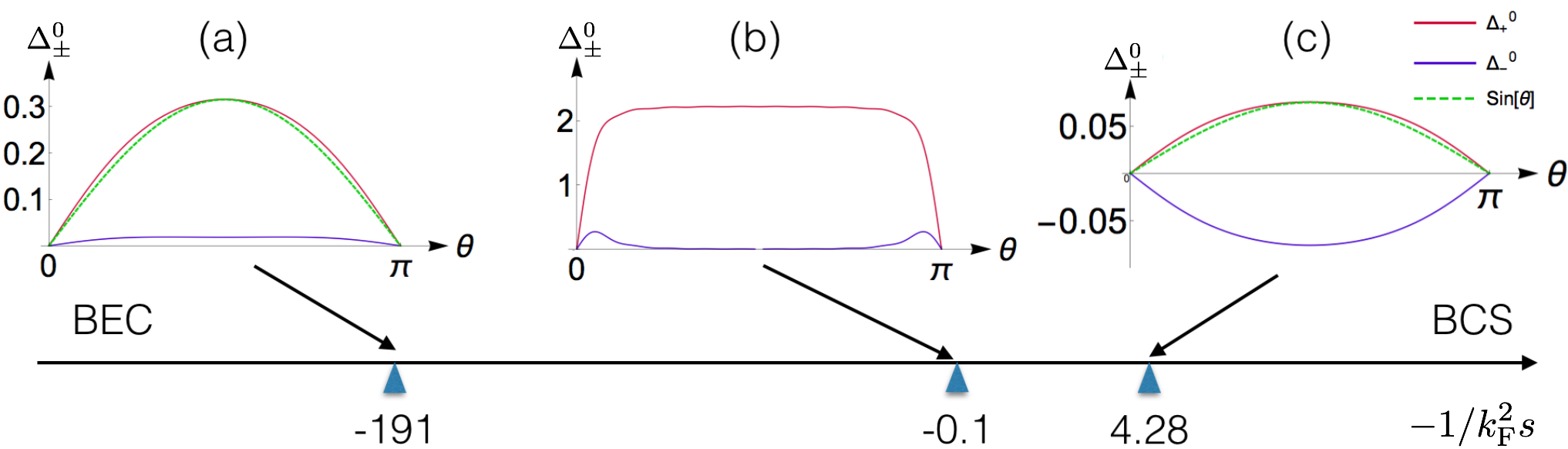}
	\caption{The order parameter $\Delta_\pm^0(\theta)$ in $m=0$ channel for different interaction strength $1/k_F^2s$. For figure (a) and (b) $1/k_FR=0.14$, for figure (c) $1/k_FR=1.38$\cite{explain}}\label{m=0}
\end{figure*}

{\it Many-body ground state.} To study the ground state of the many-body system, we apply the BCS theory. By approximating the molecule field $d_\alpha$ by its expectation value $\Delta_\alpha(x)\equiv\langle d_\alpha(x)\rangle$ (correspondingly, $\Delta_\pm(x)$ is the order parameter of $p_\theta\pm ip_\phi$ superfluid), we obtain the mean-field Hamiltonian,
\begin{eqnarray}
	\mathcal{H}_{\text{BCS}}&=&\Delta_\alpha^*\bigg(-\frac{D^\nu D_\nu}{4m_f}+\epsilon-2\mu\bigg)\Delta^\alpha\nonumber\\
	&&+\frac{1}{2}\left(
	\begin{array}{cc}
		\psi^\dagger&\psi
	\end{array}
	\right)
	\left(
	\begin{array}{cc}
		\hat{h}_0& -2i\lambda\Delta_\nu D^\nu \\
		-2i\lambda\Delta_\nu^*D^\nu & -\hat{h}_0 \\
	\end{array}
	\right)
	\left(
	\begin{array}{c}
		\psi\\
		\psi^\dagger
	\end{array}\right),\nonumber
\end{eqnarray}
where $\hat{h}_0=\frac{L^2}{2m_f}-\mu$.

The mean-field Hamiltonian is diagonalized through a Bogoliubov transformation $\alpha_i=\int \mathrm{d}s\ u_i^*(x)\psi(x)+v_i^*(x)\psi(x)^\dagger$. $(u_i, v_i)^T$ is the normalized Nambu spinor field on the sphere and satisfies the BdG equations,
\begin{eqnarray}
	\left(
	\begin{array}{cc}
		\hat{h}_0&O\\
		O^\dagger&-\hat{h}_0
	\end{array}
	\right)
	\left(
	\begin{array}{c}
		u_i\\
		v_i
	\end{array}
	\right)
	=E_i\left(
	\begin{array}{c}
		u_i\\
		v_i
	\end{array}
	\right),\label{BdG_eq}
\end{eqnarray}
where $O=-2i\lambda\Delta_\nu D^\nu-i\lambda(D^\nu \Delta_\nu)$. It is straightforward to show that $O$ satisfies $O^\dagger=-O^*$ which preserves the particle-hole symmetry.

The gap equation is obtained through the variation of the ground state energy $\delta \langle H\rangle_g/\delta \Delta_\alpha^*=0$, which gives
\begin{eqnarray}
	\bigg{(}\frac{(\mathbf{L}\pm\hat{r}\times\mathbf{A})^2}{4m_f}+\epsilon-2\mu\bigg{)}\Delta_\pm=i\lambda\sum_nu_iD_\pm v_i^*.
\end{eqnarray}
And the number equation is given by
\begin{eqnarray}
	\langle {N}\rangle=\int\mathrm{d}s\bigg{(}\langle\psi^\dagger\psi\rangle+2\Delta_\alpha^*\Delta^\alpha\bigg{)},\label{number_eq}
\end{eqnarray}
where
$\langle\psi^\dagger\psi\rangle=\sum_i|v_i|^2$. The summation $\sum_i$ should be restricted to $0\leq E_i\leq \Lambda^2/2m_f$\cite{cutoff}.

To solve the BdG equation together with the gap equation and the number equation self consistently, we expand the order parameter as
\begin{eqnarray}
	\Delta_\pm(\theta,\phi)=\sum_m\Delta^m_\pm(\theta)e^{im\phi}.
\end{eqnarray}
Note that both the BdG equation and the gap equation preserve the SO$(2)$ rotation symmetry along $z$ axis. This allows us to simplify the numerics by assuming the order parameter only contains one $m$ component. And the symmetry will assure the other $m$ components remain zero during the iterative process.

Thus, for different quantum number $m$, we may calculate the order parameter $\Delta^m_\pm(\mathbf{\theta})$ separately. And the ground state is obtained by comparing their free energies.

{\it Numerical results.} Since the effective range for $p$-wave scattering is approximately a constant across the Feshbach resonance\cite{p-wave exp1,p-wave exp2,p-wave exp3,p-wave exp4,p-wave exp5}, we fix the ratio of effective range and the radius of the sphere $r_0/R=0.01$ in all calculations. Then the ground state properties are fully controlled by two dimensionless parameters $1/k_FR$ and $-1/k_F^2s$, where  $k_F$ is related to the fermion density $n$ by $k_F=\sqrt{4\pi n}$.

Here we only present the results of $m=0$ and $m=\pm1$ channels since the numerics show that these are the three lowest energy channels. They are smooth connected to the three degenerate monopole harmonics in the BEC limit.
In the following, we first discuss these two cases separately and compare their free energy afterwards.

In Fig.(\ref{m=0}), we plot the order parameter in $m=0$ channel $\Delta_\pm^0(\theta)$ for different $-1/k_F^2s$. Firstly, we could see that the order parameters always vanish at the north and south pole. Although the order parameters are real, the emergent gauge field gives nonzero velocity\cite{Moroz}, which leads to a $2\pi$ phase winding. This is why the $m=0$ solutions correspond to a phase with two vortices locate on the north and south poles respectively. Secondly, as we tune the interaction, there are two significant phenomena demonstrating the effects of sphere from two aspects.

One is the change of vortex size. When we tune the interaction towards the BEC($-1/k_F^2s\rightarrow-\infty$) or the BCS($-1/k_F^2s\rightarrow\infty$) side, the vortices grow larger, which is consistent with the trend of healing length in a planar system. In the extreme BEC or BCS limit, the vortex size saturates when it is comparable with the system size $R$. And the order parameters are well approximated by monopole harmonics $D_{0,\pm1}^{1*}\propto\sin\theta$.
This is because in both limits the interaction terms in the Ginzburg-Landau theory are negligible and the minimization of kinetic energy forces the order parameters to keep in the lowest Landau level, whose wave function is given by the monopole harmonics.

The other is the change of the ratio between $\Delta_+$ and $\Delta_-$. In subplot (a) and (b), we can see that the order parameter $\Delta_+$ is much larger than $\Delta_-$, which means the ground state is a $p_\theta+ip_\phi$ superfluid. This is also consistent with the previous study on $p$-wave superfluid on the plane\cite{Andreev,Gurarie}. While this conclusion changes completely in the deep BCS regime. Since the Cooper pair size grows exponentially in this limit, it will soon become comparable with the radius $R$. This makes the Cooper pair feels the curvature of the sphere. As a result, the two-body process that couples $\Delta_+$ pairs to $\Delta_-$ pairs becomes very significant in BCS side. Indeed, our numerics shows that the system is in $p_\phi$ pairing instead of $p_\theta+ip_\phi$ in BCS limit.

\begin{figure}\centering
	\includegraphics[height=1.8in]
	{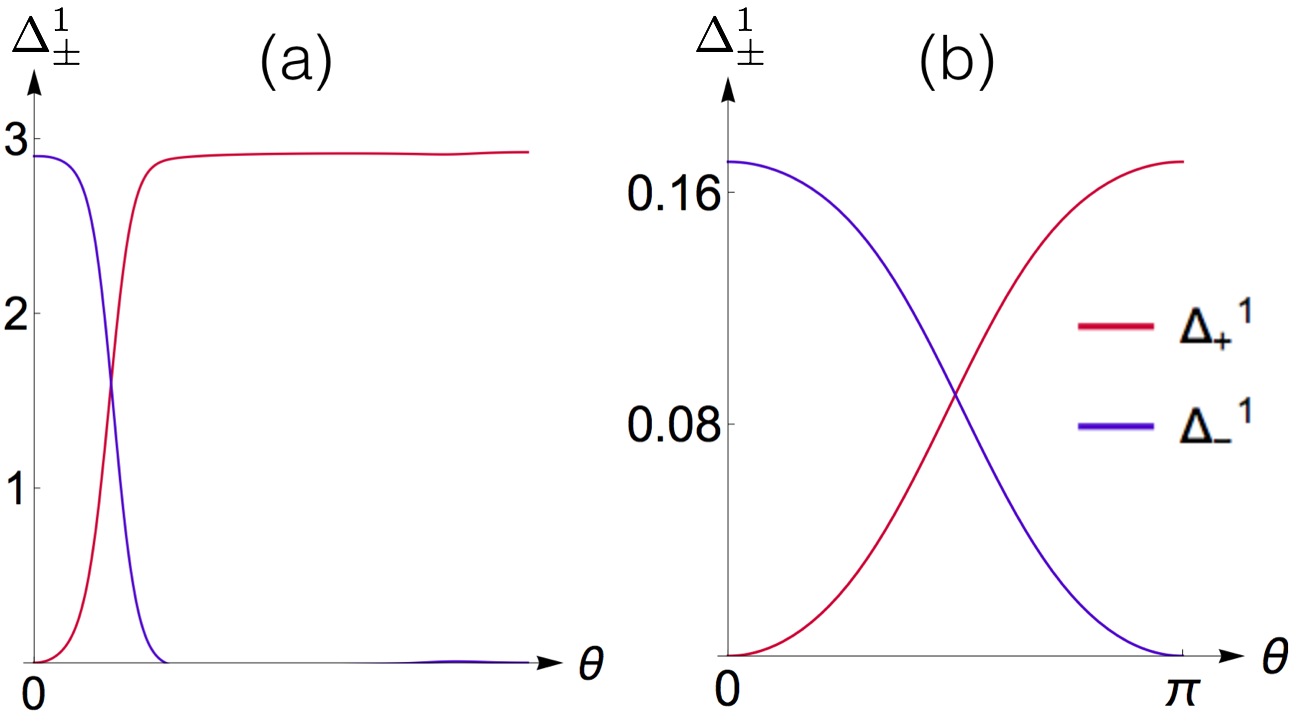}
	\caption{The order parameter $\Delta^1_\pm$ as functions of $\theta$. $1/k_\text{F}R=0.07$ (a): $-1/k_F^2s=0.76$ (b): $-1/k_F^2s=3.37$}\label{m=1}
\end{figure}
In fig.(\ref{m=1}) we plot the order parameters in $m=1$ channel $\Delta_\pm^1(\theta)$ for different interaction strength $-1/k_F^2s$ (the $m=-1$ channel are related by a reflection). Since the numerics show that the $m=1$ phase only exists in BCS side, we only present the result in $s<0$ regime. One can see that the $m=\pm1$ channel corresponds to the configuration with a domain wall between $p_\theta\pm ip_\phi$ phases.

Similar to the $m=0$ phase, the size of the domain wall also increases towards the BCS limit. In the deep BCS regime, the domain wall moves to the equator and its size saturates when it is comparable with the radius of the sphere. And the order parameters $\Delta_\pm$ are well approximated by monopole harmonics $D_{1,\pm1}^{1*}\propto(1\mp\cos\theta) e^{i\phi}$.

After comparing the energies of different $m$ channels, we obtain a phase diagram of the system. As shown in fig.(\ref{phase_d}), the ground state may exhibit topological defects such as vortex pairs($m=0$) or domain walls($m=\pm1$) depending on the interaction strength $-1/k_F^2s$ and radius $1/k_FR$. 

In the BEC limit, the system is described by repulsive interacting molecules condensing at the lowest Landau level. Then it is straightforward to show that the interaction energy of $m=0$ channel is lower. Numerics suggest this remains ture for the whole BEC side and the phase diagram on this side can be understood. 

On the BCS side, the system may undergo phase transitions from the $m=\pm1$ phase to the $m=0$ phase, when we change the radius $R$. Recall that the domain wall energy is proportional to its length. This makes the domain wall phase energetically unfavored for large $R$, which gives the phase transition in $1/k_\text{F}R \rightarrow 0$ limit. However, in $1/k_FR\rightarrow\infty$ limit, this argument breaks down since the healing length is now comparable to the system size $R$. To see what happens in this limit, we did an calculation based on Ginzburg-Landau theory (see appendix). Indeed, it shows that the vortex pair phase has lower free energy in small $R$ limit.

Furthermore, the $m=\pm1$ phase can be further separated into two phases. In (c) phase, the domain wall is exactly on the equator. While in (b) phase,the Z${}_2$ reflection symmetry is broken, and the domain wall moves to the northern or southern hemisphere. The Z${}_2$ phase transition can be explained by a  competition between the domain wall energy and the kinetic energy of the superfluid\cite{Moroz}.

{\it Conclusions.} We consider a microscopic model describing fermions with p-wave interaction on a sphere and study its ground state properties. We show that the curvature of the sphere induces an emergent magnetic monopole field coupled to the closed channel molecules, which necessarily leads to topological defects in superfluid ground state. By solving the BdG together with gap and number equations self consistently, we identify three phases with different types of topological defects and obtain the phase diagram.

{\it Acknowledgment} We are grateful to Hui Zhai, Ran Qi and Zhenhua Yu for helpful discussion. Especially we would like to thank Hui Zhai for helpful suggestions and careful reading of this manuscript.

\section{Appendix: Ginzberg-Landau in BCS limit}
The Hamiltonian $\mathcal{H}_{\text{BCS}}$ is largely simplified if we assume $\Delta_{\pm}$ to be some function with undetermined numerical factors. In $1/k_FR\rightarrow\infty$ limit, inspired by numerics, we assume $\Delta_+=-\Delta_-=\Delta \sin\theta/\lambda$ for $m=0$ channel, where $\Delta$ is a constant. Set the mass of the atom to be 1, We have:
\begin{align}
\Delta_+ \psi^\dagger i D_-\psi^\dagger+\Delta_- \psi^\dagger i D_+\psi^\dagger=\sqrt{2}\Delta \psi^\dagger \partial_\phi\psi^\dagger/\lambda
\end{align}

Hence the Hamiltonian can be written as:
\begin{align}
\int d\Omega\ \frac{1}{2}\Psi^\dagger\mathcal{H}\Psi+\frac{2\Delta^2}{\lambda^2}\sin^2\theta(\frac{1}{4}-2\mu+\epsilon)
\end{align}

$\Psi$ is the two component spinor $(\psi,\psi^\dagger)^T$ and $\mathcal{H}=(\frac{L^2}{2}-\mu)\sigma_z-i2\sqrt{2}\Delta\partial_\phi\sigma_y$. Expand the result in spherical harmonics and diagonalize the matrix give us the energy for ground state:
\begin{align}
E=&-\frac{1}{2}\sum_{lm}\sqrt{(\frac{l(l+1)}{2}-\mu)^2+8\Delta^2m^2}\notag \\ &+\frac{16\pi\Delta^2}{3\lambda^2}(\frac{1}{4}-2\mu+\epsilon)
\end{align}

Similarly, for $m=1$, the numerics suggest an assumption: $\Delta_\pm=\Delta e^{i\phi}(1\mp\cos\theta)/\sqrt{2}\lambda$. Then we have:
\begin{align}
\Delta_+ \psi^\dagger i D_-\psi^\dagger+\Delta_- \psi^\dagger i D_+\psi^\dagger=-\Delta \psi^\dagger iL_+\psi^\dagger/\lambda
\end{align}

We have used the fact that $L_{\pm}=e^{\pm i\phi}(\pm\partial_\theta+i\cot\theta\partial_\phi)$. Similar procedure give the energy:
\begin{align}
E=&-\frac{1}{2}\sum_{lm}\sqrt{(\frac{l(l+1)}{2}-\mu)^2+4\Delta^2(l-m)(l+m+1)}\notag \\ &+\frac{16\pi\Delta^2}{3\lambda^2}(\frac{1}{4}-2\mu+\epsilon)
\end{align}

In BCS limit the $\Delta$ is small. We could expand energy for $\Delta$ and get a Ginsberg-Landau free energy:
\begin{align}
E=-r\Delta^2+\frac{b}{2}\Delta^4
\end{align}

The energy have its minimum at $\Delta^2=\frac{r}{b}$ and lead to $E=-r^2/2b$. 

Using $$\sum_1^ni^2=\frac{n(n+1)(2n+1)}{6},$$ it is easy to show that the second order ($\Delta^2$) contribution from the summation is the same for any $l$ after sum over $m$. Then $r$ is the same for $m=0$ and $m=1$. For $b$, 
\begin{align}
b_{m=0}=\sum_l\frac{32  l (l+1) (2 l+1) \left(3 l^2+3 l-1\right)}{15 \left(l^2+l-2 \mu \right)^{3/2}}\\
b_{m=1}=\sum_l\frac{32  l (l+1) \left(4 l^3+6 l^2+4 l+1\right)}{15 \left(\left(l^2+l-2 \mu \right)^2\right)^{3/2}}
\end{align}

Then a subtraction tell us that:
\begin{align}
\delta b=b_{m=0}-b_{m=1}=\sum_l\frac{32 l (l+1) \left(2 l^3+3 l^2-3 l-2\right)}{15 \left(\left(l^2+l-2 \mu \right)^2\right)^{3/2}}
\end{align}

Recall that a larger $b$ gives a higher energy. For fixed density, the contribution comes mainly near $l(l+1)\sim 2\mu$, then for large density, the $m=1$ wins while for extremely small density(large $1/k_FR$), $m=0$ wins. The numerical calculation supports these results.

\begin{thebibliography}{99}
\bibitem{cutoff}
In atomic gases, the momentum cutoff is set by the inverse of the van der Waals radius, which remains a constant across Feshbach resononses. In our calulations, we also fix the cutoff $\Lambda=80/R \gg k_\text{F}$.
\bibitem{explain}
Due to the limit of the computation power, it is difficult to keep a high density to approach the BEC limit with high precision.
\end{thebibliography}
\end{document}